\documentclass [12pt] {article}
\newcommand{\nwc}{\newcommand}
\nwc{\bea} {\be\ba{rcl}}
\nwc{\eea} {\ear\ee}
\nwc{\Tr} {\rm Tr}
\nwc{\LL} {\cal L}  
\begin{document}
\begin{titlepage}
\quad\\
\vspace{1.8cm}

\begin{center}
{\Large Phase transition between three- and two-flavor QCD?}\\
\vspace{2cm}
Christof Wetterich\footnote{e-mail: C.Wetterich@thphys.uni-heidelberg.de}\\
\bigskip
Institut  f\"ur Theoretische Physik\\
Universit\"at Heidelberg\\
Philosophenweg 16, D-69120 Heidelberg\\
\vspace{3cm}

\end{center}

\begin{abstract}
We explore the possibility that QCD may undergo a phase transition
as a function of the strange quark mass. This would hint towards
models with ``spontaneous color symmetry breaking'' in the 
vacuum. For two light quark flavors we
classify possible colored
quark-antiquark, diquark and gluon condensates that are compatible
with a spectrum of integer charged states and conserved 
isospin and baryon number. The ``quark mass phase transition''
would be linked to an unusual realization
of baryon number in QCD$_2$ and could be tested
in lattice simulations. We emphasize, however, that at the
present stage the Higgs picture of the vacuum cannot predict a
quark mass phase transition - a smooth crossover remains as a
realistic alternative. Implications of the Higgs picture for
the high density phase transition in QCD$_2$ suggest
that this transition is characterized by the
spontaneous breaking of isospin for nuclear and quark matter.

\end{abstract}
\end{titlepage}

\section{Introduction}

Spontaneous color symmetry breaking by a colored quark-antiquark
condensate in the vacuum has been proposed \cite{GM}
as a ``complementary'' or ``dual'' picture for low energy QCD with three flavors of
light quarks. This idea relies
on the correspondence between a confinement and a Higgs
description \cite{CC} and finds an analogy in high density
quark matter \cite{CFL}. First dynamical investigations
suggest that the color octet $\bar qq$-condensate is induced
dominantly by instanton effects \cite{IN}, with fermion
fluctuation effects going into the same direction \cite{MF}.
The phenomenological success of this description for realistic
QCD with three flavors of light quarks \cite{GM} includes the hadronic and
leptonic decays of the $\rho, K$ and $\pi$-mesons,
including explanations of vector dominance and the $\Delta I=1/2$
rule for weak kaon decays. The coincidence of deconfinement
with the high temperature chiral phase transition in QCD results
from the melting of the octet condensate at high temperature
\cite{HT}.

The idea is intriguing enough that one may look for
possible tests to verify or falsify this picture. Unfortunately,
finding a clear-cut test is not so easy. The first is, of course,
comparison with observation. On a rough level, the model does
actually surprisingly well. Finding decisive quantitative precision
tests is hindered at present by the lack of knowledge of the effective
action relevant for low momentum scales. As long as the parameters
in the effective action are not computed from QCD there remains
substantial freedom to adapt parameters to observation, limiting the
predictivity. Second, direct tests by lattice simulations require some
thought since on a fundamental level gauge symmetries are never
spontaneously broken and the Higgs picture is only an approximate
language -- which may nevertheless be very useful, as is well known
from the electroweak gauge theory. In particular, the color octet
$\bar qq$-bilinear has a zero expectation value in any
gauge-invariant formulation. In lattice simulations a
nonzero octet expectation value could be seen only in an appropriately
gauge-fixed version. The difficulty of finding a direct simple lattice
test is actually quite profound and can be traced back to the equivalence
of the Higgs and confinement pictures. Needless to say that the
identification of a suitable test quantity and its measurement by
a simulation would be of great value.

As a third possibility, one may want to look more explicitly into
the proposal that instanton dynamics is responsible for ``spontaneous
color-symmetry breaking''. Simulations based on instanton ensembles
do not exhibit the full local color symmetry and therefore
could, in principle, show an octet condensation.  The problem with
a first simulation \cite{TSS} concerns the limited value of an instanton
ensemble where the large-scale instantons are removed ``by hand''.
These large-scale instantons are responsible
for the octet condensation in the computation of \cite{IN}. In fact,
the octet condensate is supposed to provide the infrared cutoff
for the instanton ensemble since it leads to a nonvanishing gluon mass.
In more explicit terms an instanton computation should proceed in
two steps. For the functional integral over gauge fields and quarks one
first performs the integration over gauge fields for fixed values
of the fermion fields and approximates it by a suitable integral
over instantons\footnote{The integration over the fermionic zero
modes may be included in the first step.}. The second step solves the
remaining functional integral over instanton collective
coordinates and fermions by simulation. The first step
needs the computation of instanton solutions in presence
of nonzero fermion-bilinear sources (currents) for the gauge field.
For colored bilinears this introduces an important dependence of the effective instanton
ensemble on the fermion bilinears which has not been taken into
account so far. Including the corresponding interactions for large
$\bar qq$ is crucial for the proposed IR cut-off \cite{IN} and
mandatory for a test of this idea.

Waiting for a possible test along these lines one may investigate,
as a fourth alternative, if a characteristic dependence on some
``external parameters'' like temperature or quark masses
could lead to a test. These parameters are not easily varied in
nature, but for lattice simulations this is less of a problem.
For the temperature dependence the octet condensate does
actually fine. It can explain \cite{HT} in a natural way the coincidence
of chiral symmetry restoration and deconfinement at the same critical
temperature and leads to a realistic $T_c\approx$ 170 MeV for
three light quarks with equal mass. In the present paper we
look for possible tests via the dependence of strong interaction
physics on the quark masses.

The realization of a Higgs picture of the QCD vacuum depends strongly
on the number of light quark flavors. Its original proposal relies
on the equality of the number of light flavors and the number
of colors which permits a ``color-flavor-locked'' \cite{IN}
diagonal global $SU(3)$ symmetry. On the other side, no
$\bar qq$-condensate is available for an infrared cut-off in pure
QCD without quarks (gluodynamics). If the octet condensate picture
\cite{GM} applies for three light quark flavors, one would
expect an important qualitative change as the three equal quark
masses increase. Beyond a critical quark mass the infrared
cut-off of gluodynamics is expected to dominate, and a possible
octet condensate should disappear. It is conceivable that this
qualitative change becomes visible in lattice simulations as
a phase transition or a relatively sharp crossover
as a function of the quark mass. This could be tested
once realistic quark masses can be attained in simulations with
dynamical fermions.

For two light flavors or the limit of a large strange quark
mass the situation is even more complicated.
A  simultaneous condensation of $\bar qq$-pairs
and diquarks has been proposed for the vacuum of
QCD$_2$ \cite{BerW}. The diquark condensate
introduces very interesting
features which distinguish QCD$_2$ from the
three-flavor case QCD$_3$. Notably, baryon number is spontaneously
broken by the diquark condensate and replaced by a new type
of conserved baryon charge $B'$. Whereas two of the six quarks carry
$B'=1$ and can be identified with the proton and neutron, the other
four have $B'=0$. This leads to an interesting Higgs description
where quarks carry fractional quark-baryon charge $B_q=1/3$ whereas
they are integer charged with respect to $B'$. This particular
feature of QCD$_2$ influences strongly the transition to the high
density and high temperature state \cite{BerW},\cite{BerHD}.

Lowering the strange quark mass from high values to zero moves
us from QCD$_2$ to QCD$_3$. The different realization of baryon number
in QCD$_2$ and QCD$_3$ suggests the possibility that a phase transition could occur
at some critical strange quark mass $m_{s,c}$. For this critical value the baryon
charge of the hyperons $\Sigma$ and $\Lambda$ switches from $B'=0$ (for
QCD$_2$) to $B=1$ (for QCD$_3$). A similar change occurs for the
effective strangeness quantum number. From the observed realization
of $B$ and $S$ one concludes that $m_{s,c}$ should be higher
than the physical value of $m_s$. A phase transition
as a function of $m_s$ can be studied by lattice simulations.
Since the existence and the details of such a phase transition depend crucially
on the condensates in the vacuum of QCD$_2$, in particular
on the occurrence of diquark condensation, the Higgs picture of the QCD vacuum
cannot clearly predict the presence of a transition at the present stage.
Nevertheless, the observation of such a transition in lattice
simulations would give a very important hint about the vacuum condensates in QCD!

In fact, in absence of colored condensates one expects that the usual
color-singlet $\bar qq$-condensates change smoothly as a function
of the quark mass. At least there is no obvious reason for a phase
transition. The observation of a phase transition between QCD$_3$ and QCD$_2$ as a function
of $m_s$ or between QCD with three or two
light flavors and gluodynamics could therefore be interpreted as a
rather clear signal for the occurrence of other condensates. In particular, we will see
that it could find a natural explanation within a Higgs picture
with an octet-condensation in the vacuum!

In this note we explore the possibility of testing the Higgs
picture of the QCD vacuum by a phase transition as a function
of $m_s$ in more detail. For this purpose
we present a systematic study of possible symmetry
breaking patterns in QCD$_2$ with two flavors.
We restrict our discussion
to vacuum states which preserve a global isospin symmetry such that
the eight gluons transform as a triplet $(\rho)$ two
doublets $(K^*,\bar K^*)$ and a singlet $(\omega)$. Similarly,
two of the quarks should carry the quantum numbers of the proton
and the neutron. Our classification of $\bar qq$, diquark and
gluonic condensates carries over to the high density state of
QCD$_2$. Dynamical arguments point to a simultaneous condensation of
$\bar qq$ and diquarks in the high density state of QCD$_2$ \cite{BerHD}.
This implements
color-flavor locking for two flavors.
Our discussion of arbitrary expectation values of isospin-conserving
$\bar qq$- and diquark operators partly overlaps with other
discussions of high density condensates \cite{HDS, CFL}.
For the vacuum state we extend the discussion of \cite{BerW}
by the inclusion of other possible condensates\footnote{See also
\cite{OGT} for early discussions of the Higgs mechanism for $SU(3)$-gauge theories with fundamental scalars.}.
Since all arguments presented here
are solely based on symmetry properties they apply independently
of a given dynamical scenario which would explain
how the color symmetry breaking condensates are generated.

Besides the possible
phase transition between QCD$_3$ and QCD$_2$ as a function of the strange
quark mass we are interested here in a second  issue. We want to
classify the characteristics of a possible
high density phase transition in QCD$_2$ in
dependence on the baryon density, assuming\footnote{Without a
colored vacuum condensate this phase transition is extensively
discussed in \cite{HDS}.} that a colored
condensate occurs in the vacuum.
For both purposes we first study possible condensates
which might be relevant for the ground state or the high density
state of QCD$_2$. We begin in sect. 2 with quark-antiquark condensates
transforming as color octets. This discussion is extended in sect. 3
to include diquark condensates whereas gluonic condensates in nontrivial color
representations are added in sect. 4. 

In sect. 5 we turn to the transition between QCD$_2$ and QCD$_3$ as the
strange quark mass is lowered from very large values towards zero.
If QCD$_2$ is characterized by a diquark condensate, a phase transition
is plausible (but not necessary). The high density phase transition
in QCD$_2$ is discussed in sect. 6. Its characteristics depend on
the condensates in the vacuum, in particular on the question of
a vacuum-diquark condensate. A continuous crossover to
the high density phase becomes possible if isospin is conserved
in the high density phase. We suggest that spontaneous breaking
of isospin in the high density phase leads to a true phase
transition. The spectrum of excitations in nuclear and quark
matter contains in this case a massless Goldstone boson $a^o$ responsible for superfluidity
as well as two light pseudo-Goldstone bosons $a^\pm$ in addition to the
pions.
Furthermore, we present in sect. 7 a first only partially successful attempt to
understand the ground state of pure QCD (no light quarks) in terms of
color symmetry breaking gluonic condensates. Finally, in sect. 8 we discuss the transition
to gluodynamics when all three light quarks get heavy simulaneously. In particular, we
investigate the consequences of the dual description of the QCD-vacuum by a Higgs
and confinement picture for the shape of the heavy quark potential. Our conclusions
are presented in sect. 9.

Our main findings are:
\begin{enumerate}
\item For QCD$_2$ the condensation of both a color
octet quark-antiquark pair and
an antitriplet (or sextet) diquark pair in the vacuum would lead to
complete ``spontaneous symmetry breaking'' for the local color
group \cite{BerW}. All gluons acquire a mass. A residual global $SU(2)$
isospin symmetry and global baryon number $B'$ for the proton and
neutron remain preserved. The local abelian electromagnetic
symmetry leads to integer charges $Q$ for the quarks. In the Higgs
picture the quarks are identified with baryons. Chiral
symmetry is spontaneously broken and the baryons are massive. The massive gluons are integer charged and can be associated with eight vector
mesons. Without the diquark condensate one of the gluons corresponding
to the $\omega$-meson would remain massless.

\item As the strange quark is added and its mass lowered, two flavor
QCD transmutes into realistic QCD and finally into three-flavor QCD
with three massless or very light quarks. For a very heavy strange
quark the dynamics remains essentially the same as for two-flavor
QCD. A new global symmetry is added, however, which acts on the
new massive excitations. In the vacuum it corresponds to a new
conserved strangeness quantum number $S'$. All $s$-quarks have $S'=-1$.
The QCD vacuum with diquark condensation
exhibits now three independent global symmetries
(in absence of electromagnetism) which correspond to $Q, B',$ and $S'$.

\item The standard symmetries $B$ and $S$ are not realized
by the vacuum or high density state of QCD$_2$ if
all gluons acquire a mass from isospin-conserving
scalar condensates. These generators are broken by
the diquark condensate. The spontaneous breaking of $B$ (or $S$)
induces, however, no Goldstone boson since it is accompanied
by the breaking of a local symmetry which is part of the
color symmetry. The ``would-be Goldstone
boson'' is eaten by the Higgs mechanism. (It is the longitudinal
component of the $\lambda_8$-gluon or $\omega$-meson.)

\item For a light strange quark mass equal to the up and down  quark mass
one expects the unbroken global
$SU(3)$-symmetry of QCD$_3$. In this case the color octet
condensate gives mass to all gluons. In particular, the $\omega$ acquires a
mass from the condensation of the strangeness components of the color
octet quark-antiquark pair. No diquark condensate is possible in the vacuum
since this would break baryon number. The preserved exact global
symmetries correspond now to $Q, B$ and $S$. The quantum numbers $B$
and $S$ for the fermions in QCD$_3$ differ from $B'$ and $S'$ for QCD$_2$
in presence of a diquark condensate. In this case one
therefore expects a transition as a function of the strange quark
mass $m_s$, from a diquark condensate for large $m_s$ to an
(almost) $SU(3)$-symmetric octet condensate for small $m_s$.

\item The transition between QCD$_3$ and QCD$_2$ may be associated with
a phase transition (as a function of $m_s$). We note that an
intermediate state with both an antitriplet diquark condensate
and nonvanishing strangeness components of the octet condensate would
lead to a state with a different realization of the symmetries. One of
the global symmetries $(S)$ is broken in such a state, implying the existence
of an exactly massless Goldstone boson and therefore superfluidity.
The transition between large and small $m_s$ could therefore  either
proceed directly by a first-order phase transition or pass by
two transitions with an intermediate superfluid phase. As an alternative
one may conceive
an intermediate Coulomb-like phase where only the nonstrange
components of the octet condense and one gluon remains
massless in the Higgs picture. As a general remark we recall, however,
that the Higgs picture is only an approximation and may not be reliable for
all questions. Since no global symmetries of the vacuum are altered as
$m_s$ switches from large to small values an analytic crossover remains possible as well.

\item We suggest that the high density transition to nuclear matter is
characterized by a spontaneous
breaking of the global isospin symmetry due to a di-neutron
condensate. In this event an exactly massless Goldstone boson $a^0$
exists in nuclear matter, corresponding to the spontaneous breaking of
the global $I_3$-symmetry. (The latter is an exact
symmetry even in presence
of nonvanishing quark masses and electromagnetism.) In case of spontaneous isospin breaking,
nuclear matter is a superfluid. In addition, two new light scalars
$a^\pm$ correspond to the breaking of the generators $I^\pm$. The
mass of these pseudo-Goldstone bosons vanishes in
the limit of equal up and down
quark masses and absence of electromagnetism.
This behavior is similar to the well-known
pions. Spontaneous isospin breaking may occur also
for the high density ``quark matter'' phase of QCD$_2$.

\end{enumerate}

Before starting with a more detailed description of the different
possible ``states'' of QCD$_2$ in the main part of this paper, we present
a brief summary in table 1. Here the different possible condensates
are denoted by their $SU(3)_c$-representation with the $SU(2)$-flavor
representation as a subscript. The naming and details are explained in the
main text. We do not list the usual color singlet $\bar qq$-condensate
which may always accompany the other condensates. The standard
picture of the QCD vacuum would have vanishing expectation values
for all condensates listed in table 1.

\medskip
\begin{center}
\begin{tabular}{|l|c|c|c|c|c|}
\hline
&\multicolumn{2}{|c|}{condensate}&
superfluidity&electric charge&massless\ gluons\\
\hline
A&$8_3$&$\bar\xi_1$&no&integer&1\\
B&$8_3+\bar 3_1$&$\bar\xi_1,\bar\delta_1$&no&integer&0\\
C&$\bar 3_1$&$\bar\delta_1$&no&integer&3\\
\hline
D&$8_3,8_2$&$\bar\xi_1,\bar\xi_2$&no&integer&0\\
\hline
E&$8_3, 8_2,\bar 3_1,$&$\bar\xi_1,\bar\xi_2,\bar\delta_1$&yes&integer&0\\
\hline
F&$8_3,10$&$\bar\xi_1,\bar a$&no&fractional&0\\
G&$8_3,\bar 3_1,10$&$\bar\xi_1,\bar\delta_1,\bar a$&no&
fractional&0\\
\hline
\end{tabular}\\

\medskip Table 1: States with conserved $SU(2)_I$
\end{center}

\section{Quark-antiquark color octet}

\medskip

Let us start with a discussion of the color-symmetry-breaking
pattern induced by quark-antiquark condensates in the vacuum of QCD$_2$.
This corresponds to the entry (A) in table 1.
With respect to the $SU(3)_c$ group a quark-antiquark pair
can be in a singlet $\tilde\varphi$ or an octet $\tilde\chi$
\begin{eqnarray}\label{A}
\tilde\varphi^{(1)}_{ab}&=&\bar\psi_{L\ ib}\ \psi_{R\ ai}\quad,\quad
\tilde\varphi^{(2)}_{ab}=-\bar\psi_{R\ ib}\ \psi_{L\ ai}
\nonumber\\
\tilde\chi^{(1)}_{ij,ab}&=&\bar\psi_{L\ jb}\ \psi_{R\ ai}
-\frac{1}{3}\bar\psi_{L\ kb}\ \psi_{R\ ak}\ \delta_{ij}\nonumber\\
\tilde\chi^{(2)}_{ij,ab}&=&-\bar\psi_{R\ jb}\ \psi_{L\ ai}
+\frac{1}{3}\bar\psi_{R\ kb}\ \psi_{L\ ak}\ \delta_{ij}\end{eqnarray}
Here the color indices $i,j$ run from 1 to 3 and the flavor indices
$a,b=1,2$ denote the up and down quark. We consider vacuum expectation
values of $\tilde\varphi$ and $\tilde\chi$ that lead to color-flavor
locking where a diagonal $SU(2)_I$-subgroup
of local $SU(2)_c$-color and global vector-like $SU(2)_F$-flavor
remains unbroken. The unbroken global ``physical'' $SU(2)_I$ symmetry
is associated with isospin. The decomposition of the $SU(2)_c\times SU(2)_F$
representations $\tilde\varphi,\tilde\chi$ with respect to
$SU(2)_I$
\begin{eqnarray}\label{B}
&&\tilde\varphi:(1,1+3)\to 1+3\nonumber\\
&&\tilde\chi:(8, 1+3)\to 1+1+2+2+2+2+3+3+3+4+4+5\end{eqnarray}
shows the existence of three singlets which can acquire
a vacuum expectation value.

We will use a bosonized language where the quark-antiquark bilinears
are replaced by scalar fields $\tilde \varphi^{(1)}_{ab}\to
\sigma_{ab},\ \tilde\varphi^{(2)}_{ab}\to \sigma^\dagger_{ab},\ \tilde \chi^{(1)}_{ij,ab}\to
\xi_{ij,ab},\ \tilde\chi^{(2)}_{ij,ab}\to\xi^*_{ji,ba}$.
The most general isospin conserving expectation values are
\begin{eqnarray}\label{C}
\sigma_{ab}&=&\bar\sigma\delta_{ab},\\
\xi_{ij,ab}&=&\frac{1}{2\sqrt6}\bar\xi_1(\lambda^k)_{ij}(\tau^k)_{ba}
+\frac{1}{6\sqrt2}\bar\xi_2(\lambda^8)_{ij}\delta_{ab}\nonumber\end{eqnarray}
where $\bar\sigma,\bar\xi_1$ and $\bar\xi_2$ correspond to the three
singlets in eq. (\ref{B}). Here $\tau^k$ are the Pauli matrices and
$\lambda^z$ the Gell-Mann matrices, with a sum $k=1...3$. We note
the presence of two color breaking directions $\bar\xi_1,\bar\xi_2$.

The physical electric charge $Q$ is composed of the quark charge
$\tilde Q(\tilde Q=-2/3$ for up, $\tilde Q=1/3$ for down) and the color
generator $Q_c$
\begin{equation}\label{D}
Q=\tilde Q-Q_c=\tilde Q-\frac{1}{2}\lambda_3-
\frac{1}{2\sqrt3}\lambda_8\end{equation}
It is easy to check that the quark and gluons have integer electric
charge. We can also relate the physical electric charge to isospin
$(I_3)$ and interpret the color $\lambda_8$-generator in terms of standard
baryon number $B$ and strangeness $S$
\begin{equation}\label{E}
Q=I_3+\frac{1}{6}B-\frac{1}{2\sqrt3}
\lambda_8=I_3+\frac{1}{2}(B+S)=I_3+\frac{1}{2}B'\end{equation}
Here we use a normalization where the baryon number\footnote{For
a detailed discussion of the issue of baryon number see \cite{GM}. The quark baryon number $B_q$ obeys $B_q=B/3$.} of the fermion fields is $B=+1$ and the ``hyperons'' $\Sigma$ and $\Lambda$ carry $S=-1$.
With respect to $SU(2)_I, S, Q$ and $B$ the six
quarks carry the quantum numbers of the baryons $(p,n,\Lambda,\Sigma)$
(see table 2). We note that two-flavor QCD has two independent
exact global symmetries in presence of nonvanishing quark masses
and absence of electromagnetism. These are $B'=B+S$ and $Q$.
In presence of electromagnetic interactions the generator $Q$ will be
gauged.

\medskip
\begin{center}
\begin{tabular}{|l|rrr|crc|cc|c|c|}
\hline
&$\tilde Q$&$ Q_c$&$Q$&$I_3$&$S$&$B$&$B'$&$S'$&$Q'$&\\
\hline
$u_1$&2/3&2/3&0&0&-1&1&0&0&1/6&$\Sigma^0,\Lambda^0$\\
$u_2$&2/3&-1/3&1&1&-1&1&0&0&7/6&$\Sigma^+$\\
$u_3$&2/3&-1/3&1&1/2&0&1&1&0&2/3&$p$\\
\hline
$d_1$&-1/3&2/3&-1&-1&-1&1&0&0&-5/6&$\Sigma^-$\\
$d_2$&-1/3&-1/3&0&0&-1&1&0&0&1/6&$\Sigma^0,\Lambda^0$\\
$d_3$&-1/3&-1/3&0&-1/2&0&1&1&0&-1/3&$n$\\
\hline
\end{tabular}\\

\medskip Table 2: Charges of the up and down quarks
\end{center}

\medskip The surprising  fact that baryons with strangeness
$S=-1$ (i.e. $\Lambda,\Sigma$) are described by up and down quarks
arises from the fact that the strangeness of the baryons has a color
component according to (\ref{E}). Actually, for the symmetry-breaking
pattern that we describe here the isospin symmetry
corresponds to a subgroup
of the physical $SU(3)$ symmetry discussed for the
three-flavor case in \cite{GM}. As compared to
the three flavor case it is sufficient to leave out
the states corresponding to the strange quark. Those
correspond to the baryons $\Xi$ and a baryon
singlet $S$ (see table 4 below).

The gluons carry the quantum numbers of the $\rho, K^*$ and $\omega$
mesons. Again, they describe also states with strangeness.
Their quantum numbers are displayed in table 3.

\medskip
\begin{center}
\begin{tabular}{|c|ccc|c|ccc|c|l|}
\hline
&$\tilde Q$&$ Q_c$&$Q$&$I$&$I_3$&$S$&$B$&$B'$&\\
\hline
$A_3$&0&0&0&1&0&0&0&0&$\rho^0$\\
$A_1+iA_2$&0&-1&1&1&1&0&0&0&$\rho^+$\\
$A_1-iA_2$&0&1&-1&1&-1&0&0&0&$\rho^-$\\
\hline
$A_4+iA_5$&0&-1&1&1/2&1/2&1&0&1&$K^{*+}$\\
$A_4-iA_5$&0&1&-1&1/2&-1/2&-1&0&-1&$K^{*-}$\\
$A_6+iA_7$&0&0&0&1/2&-1/2&1&0&1&$K^{*0}$\\
$A_6-iA_7$&0&0&0&1/2&1/2&-1&0&-1&$\overline K^{*0}$\\
\hline
$A_8$&0&0&0&0&0&0&0&0&$\omega$\\
\hline
\end{tabular}\\

\medskip Table 3: Charges of the gluons
\end{center}

As a consequence of the Higgs mechanism seven out of the eight gluons
acquire a mass. This can be inferred by inserting the vacuum
expectation values (\ref{C}) in the covariant derivative for $\xi$,
\begin{eqnarray}\label{GA}
{\cal L}_{kin,\xi}&=&\hat Z(D^\mu\xi)_{ij,ab}^*(D_\mu\xi)_{ij,ab},\nonumber\\
(D_\mu\xi)_{ij,ab}&=&\partial_\mu\xi_{ij,ab}-ig A_{ik,\mu}\xi_{kj,ab}
+ig\xi_{ik,ab}A_{kj,\mu}\nonumber\\
A_{ij,\mu}&=&\frac{1}{2} A^z_\mu(\lambda_z)_{ij}\end{eqnarray}
With $(\lambda^k)_{ij}(\tau^k)_{ba}=2\delta_{ia}
\delta_{jb}-\delta_{ij}\delta_{ab}$ we find
\begin{eqnarray}\label{F}
M^2_\rho&=&\frac{2}{3}\hat Z g^2|\bar\xi_1|^2\nonumber\\
M^2_{K^*}&=&\hat Zg^2(\frac{1}{4}|\bar\xi_1|^2+\frac{1}{12}
|\bar\xi_2|^2)\nonumber\\
M^2_\omega&=&0\end{eqnarray}
where $M_\rho$ denotes the mass of the $\rho$-triplet, $M_{K^*}$ the
one of the two isospin doublets $K^*,\bar K^*$ and $M_\omega$
concerns the isospin singlet $\omega$.
In addition to global isospin symmetry the expectation values
$\bar\xi_1,\bar\xi_2$ leave a local $U(1)_8$-subgroup of
color unbroken which corresponds to the generator $\lambda_8$.
This is the reason for the massless
$\omega$-meson. In the language of hadrons we may associate
the corresponding charge with a linear combination of strangeness
$S$ and baryon number $B$,
\begin{equation}\label{F1}
\lambda_8=-\sqrt3 S-\frac{2}{\sqrt3}B\end{equation}

The unbroken local abelian symmetry is a direct consequence
of the group structure and conserved isospin. All possible vacuum
expectation values of quark-antiquark operators preserve the local $U(1)_8$
symmetry, which is a remnant of the color symmetry. Independent
of the detailed dynamics one therefore expects for a state
were only octets condense a massless gauge boson similar to the photon,
but with a strong gauge coupling. It is conceivable that this symmetry
is realized in the Coulomb phase. If true, the existence of a massless
spin one state could be checked by a numerical simulation of
QCD$_2$. We note that $\bar\xi_1$
leaves the additional $U(1)_8$-symmetry unbroken and therefore
does not contribute to $M_\omega$, whereas an unbroken gauged
$SU(2)_c\times U(1)_c$ symmetry for $\bar\xi_1=0,\bar\xi_2\not=0$ forbids
a contribution of $\bar\xi_2$ to $M_\rho$ and $M_\omega$. All masses
are proportional to the strong gauge coupling $g$, with
proportionality factor associated to the wave function renormalization
$\hat Z$.

The expectation values $\bar\sigma,\bar\xi_1$ and $\bar\xi_2$
spontaneously break the chiral $SU(2)_L\times SU(2)_R$
symmetry, resulting in three (almost) massless pions. (There are no
pseudo-Goldstone bosons transforming as kaons in the two flavor
picture.) Chiral symmetry breaking also gives a mass to the
fermions which we associate with the baryon masses. We write the
relevant Yukawa-type interaction in the form
\begin{eqnarray}\label{G}
{\cal L}_y&=&\bar\psi_{ia}(h_\sigma\sigma_{ab}\delta_{ij}+h_\xi\xi_{ij,ab})
\frac
{1+\gamma_5}{2}\psi_{bj}\nonumber\\
&&-\bar\psi_{ia}(h_\sigma\sigma^\dagger_{ab}\delta_{ij}+
h_\xi\xi^*_{ji,ba})\frac{
1-\gamma_5}{2}\psi_{bj}\end{eqnarray}
The corresponding contribution to the masses of the nucleons and
hyperons $(\Sigma^0=\frac{1}{\sqrt2}(u_1-d_2),
\ \Lambda^0=\frac{1}{\sqrt2}(u_1+d_2))$ are
\begin{eqnarray}\label{9A}
M_n&=&h_\sigma\bar\sigma-\frac{1}{3\sqrt6}h_\xi\bar\xi_2\nonumber\\
M_\Sigma&=&h_\sigma\bar\sigma-\frac{1}{2\sqrt6}h_\xi\bar\xi_1+\frac{1}{6\sqrt6}h_\xi
\bar\xi_2\nonumber\\
M_\Lambda&=&h_\sigma\bar\sigma+\frac{3}{2\sqrt6}h_\xi\bar\xi_1+
\frac{1}{6\sqrt6}h_\xi\bar\xi_2\end{eqnarray}

If electromagnetic interactions are added to  QCD,
a linear combination $I_{3F}+\frac{1}{6}B$ of the previously
global symmetries becomes gauged. As a consequence, color symmetry breaking
by the octet condensate $\xi$ leaves now two gauge bosons massless,
corresponding to the two directions $I_{3F}+\frac{1}{6}B-I_{3c}$ and $\lambda_{8,c}$.
A linear combination of them is the photon.

The appearance of the massless gluon of the unbroken $U(1)_8$-symmetry
is special for the quark-antiquark condensates in QCD$_2$.
This phenomenon does not happen for three light flavors. For $N_f=2$ is
is connected to conserved strangeness and the fact that strangeness
has only a contribution from the color generator $\lambda_8$ and baryon number.
In contrast, for $N_f=3$ the strangeness
quantum number also receives a contribution
form a flavor generator and ``strangeness locking'' permits
the breaking of the local $U(1)_8$-symmetry. As long as we only
consider quark-antiquark condensates, there is no way  in two-flavor
QCD of conserving isospin in the vacuum and giving mass to the
$U(1)_8$-gluon. For $N_f=2$ we still have, however, the possibility
of condensation of diquarks or
pure gluonic operators consistent with the
physical isospin symmetry (see below). Again, this is different from three
flavors. For $N_f=3$ color-flavor locking implies that the physical
$SU(3)$-representation for the gluon degrees of freedom coincides
with the $SU(3)$-representation. All gluonic $SU(3)$-singlets
must also be color $SU(3)_c$-singlets. Singlet operators of the type $F^{\mu\nu}_{ij}\ F_{ji,\mu\nu}$ do not affect the symmetry-breaking
pattern. We conclude that despite many similarities between QCD$_2$
and QCD$_3$ the details of spontaneous color symmetry breaking
depend critically on the number of light quarks.

We finally demonstrate the connection between integer electric charge
and a massless $\omega$-meson for QCD$_2$ without diquark condensates
by a simple group-theoretical argument. The
color generator $\lambda_8$ can be represented as a
combination of isospin, baryon number and electric charge
\begin{equation}\label{Y1}
\lambda_8=2\sqrt3(I_3+\frac{1}{6}B-Q)\end{equation}
Quark-antiquark as well as gluonic operators conserve baryon number.
This implies that all
gluonic operators which conserve isospin and break the $U(1)_8$-symmetry
necessarily violate electric charge. Indeed, the 10-dimensional
$SU(3)_c$-representation (contained in the antisymmetric product of
two octets) has a $SU(2)$-singlet which is charged with respect to
$U(1)_8$ (see sect. 4). An expectation value of this
singlet would give a mass
to the $U(1)_8$-gluon. In presence of the electromagnetic gauge
interaction, however, such a vacuum would result in a modified
charge seen by the photon, given by $Q'=\tilde Q-\frac{1}{2}\lambda_3$.
(This corresponds to eq. (\ref{D}) without the piece from $\lambda_8$.)
In consequence, the physical fermions would carry electric
charges $(1/6,7/6,2/3,-5/6,1/6,-1/3)$ (cf. table 2).
Our scenario for QCD$_2$ leads
to three interesting alternatives: if isospin and $B$
are conserved in the
vacuum and all condensates are scalars, either
diquark condensation occurs or there is a massless gluon
or the electric charges
of the baryons are unusual. These alternatives
can be tested by lattice simulations.
The perhaps favored alternative is the spontaneous
breaking of baryon number in
the vacuum by a diquark condensate, which we will discuss next.

\section{Diquark condensates}

\medskip
We next turn to possible diquark condensates. They are expected to
play a role at high baryon density.
In contrast to QCD$_3$ they can also be relevant
for the vacuum in QCD$_2$ \cite{BerW}. The simultaneous octet
and diquark condensation corresponds to the entry (B) in table 1,
whereas a pure diquark condensation corresponds to (C).
Scalar diquarks are in the
antisymmetric product of two quark fields. With respect to
$SU(3)_c\times SU(2)_F$ they transform as
$(\bar 3,1)+(6,3)$. We first consider the antitriplet
\begin{equation}\label{H}
(\tilde\delta_{L,R})_i=(\psi_{L,R})_{aj\beta} c^{\beta\gamma}
(\psi_{L,R})_{bk\gamma}\epsilon_{ijk}\epsilon_{ab}\end{equation}
with $\beta,\gamma$ spinor indices and $c^{\beta\gamma}$ the
antisymmetric charge
conjugation matrix. The corresponding scalar fields $\delta_{L,R}\sim
\tilde\delta_{L,R}$
transform as a $\bar 3$ under color and a singlet under flavor.
Being flavor singlets, the expectation values of $\delta_{L,R}$ cannot
contribute to the breaking of the global $SU(2)_F$-flavor
symmetry or the chiral symmetry $SU(2)_L\times SU(2)_R$. Unlike the
case for three flavors, they cannot induce color flavor locking
for two flavor QCD. Nevertheless, they can contribute to the
breaking of color. In presence of octet condensates
$\bar\xi_1,\bar\xi_2$ the expectation value of $\delta_{L,R}$
presumably favors the isospin-conserving direction\footnote{We note the possibility
of an isospin violating alignment of the expectation values
of $\delta_{L,R}$ which breaks color completely. The orthogonal
expectation values $\delta_{Li}=\bar\delta_L\delta_{i1}, \delta_{Ri}
=\bar\delta_R\delta_{i2}$ induces a gluon mass matrix $M^2_{yz}
=Z_\delta g^2\bar\delta^2(\frac{4}{3}\delta_{yz}+\frac{2}{\sqrt3}
d_{yz8})$
or
$
M^2_\rho=2Z_\delta g^2\bar\delta^2,\ M^2_{K^*}=Z_\delta g^2\bar\delta^2,
M^2_\omega=\frac{2}{3}Z_\delta g^2\bar\delta^2.$
All gluons are massive, with an order $M_\rho^2>M_{K^*}^2>M^2_\omega$
reversed as compared to the realistic QCD vacuum. For this
alignment the standard parity transformation $\delta_L\leftrightarrow
\delta_R$ is spontaneously broken and one has to investigate if
another parity like discrete symmetry survives.}.
There is one isospin singlet for both $\delta_L$ and $\delta_R$
and parity is conserved for
\begin{equation}\label{I}
\delta_{Li}=\delta_{Ri}=\bar\delta_1\delta_{i3}\end{equation}
In addition to the global $SU(2)_F$ symmetry this expectation
value preserves a local $SU(2)_c$ subgroup of color. Therefore
$\bar\delta\not=0$ contributes to the mass of the $K^*$- and
$\omega$-vector mesons, but not to the $\rho$-mesons. From the
covariant kinetic term
\begin{eqnarray}\label{11A}
{\cal L}_{kin,\delta}&=&Z_\delta\{(D^\mu\delta_L)^*_i(D_\mu\delta_L)_i
+(D^\mu\delta_R)^*_i(D_\mu\delta_R)_i\}
\nonumber\\
(D_\mu\delta)_i&=&\partial_\mu\delta_i+ig\delta_jA_{ji,\mu}\end{eqnarray}
one finds
\begin{eqnarray}\label{J}
M_\rho^2&=&0\nonumber\\
M_{K^*}^2&=&\frac{1}{2}Z_\delta g^2|\bar\delta_1|^2\nonumber\\
M_\omega^2&=&\frac{2}{3}Z_\delta g^2|\bar\delta_1|^2\end{eqnarray}

The expectation value (\ref{I}) induces a spontaneous breaking of
baryon number. It also breaks the local $U(1)_8$-symmetry
which is preserved by $\bar\xi_1$ and $\bar\xi_2$.
In consequence the $\omega$-meson related to
the $\lambda_8$-generator of $SU(3)_c$ becomes massive
in a situation where $\bar\xi_1\not=0,\bar\xi_2\not=0,\bar
\delta\not=0$. In this setting all gluons have acquired a mass. This is
easily seen by adding the contributions (\ref{F}) and (\ref{J}).
A diquark condensate in the vacuum of QCD$_2$ is possible since
it preserves a new color-flavor-locked baryon number
\begin{equation}\label{12A}
B'=B_q-\frac{1}{\sqrt3}\lambda_8\end{equation}
with $B_q=B/3$ the quark baryon number. In terms of the hadronic
quantum numbers $B$ and $S$ this reads
\begin{equation}\label{N}
B'=B+S\end{equation}
In consequence, switching on the expectation value $\bar\delta$
preserves the total number of conserved global abelian symmetries. It
shifts the global charge from $B$ to $B+S$ and breaks the local
$U(1)_8$ symmetry. We conclude that the formation of a $\bar\delta$
condensate does not lead to a massless Goldstone boson and to superfluidity.
Regions in the QCD-phase diagram with $\bar\delta=0$ and $\bar\delta
\not=0$ can be analytically connected. The formation of a
$\bar\delta$-condensate looks similar to the phase transition in
the abelian Higgs model or to the onset of superconductivity by the
formation of Cooper pairs. Only the photon is replaced by the
$\omega$-meson. If, in addition, one gauges the electromagnetic
$U(1)$ symmetry, the true photon remains massless since the
$\delta$-condensate carries zero electric charge. Only one of the
two massless gauge bosons for $\bar\delta=0$ gets a mass for
$\bar\delta\not=0$.

The diquark contribution to the baryon masses follows from the Yukawa
coupling
\begin{equation}\label{16A}
{\cal L}=h_\delta\epsilon_{ijk}\epsilon_{ab}(\delta_{Li}^*\psi_{Laj}
c\psi_{Lbk}+L\leftrightarrow R+c.c.)\end{equation}
This yields a Majorana-type mass term for the hyperons $\Sigma$ and $
\Lambda$
\begin{equation}\label{16B}
{\cal L}=h_\delta\bar\delta_1\{
\Lambda^0_Lc\Lambda^0_L-(\Sigma_L^+c\Sigma_L^-+\Sigma^-_Lc\Sigma^+_L+
\Sigma^0_Lc\Sigma^0_L)+L\leftrightarrow R+c.c.\}\end{equation}

We next turn to the possible condensation of the color
sextet diquark. The sextet diquark is symmetric in both
color and flavor indices
\begin{equation}\label{P}
(\tilde\beta_{L,R})_{ij,ab}=\frac{1}{\sqrt2}\left\{(\psi_{L,R})_{ai\beta}
c^{\beta\gamma}(\psi_{L,R})_{bj\gamma}
+(\psi_{L,R})_{aj\beta}c^{\beta\gamma}(\psi_{L,R})_{bi
\gamma}\right\}\end{equation}
The associated scalar field $\beta_{ij,ab}^{(L,R)}$ carries again two
units of baryon number $B=2$ and transforms under $SU(2)_I$
as $1+2+3+3+4+5$. It therefore contains one singlet
\begin{equation}\label{Q}
\beta^{(L)}_{ij,ab}=\beta^{(R)}_{ij,ab}=\frac{1}{2\sqrt2}\bar\beta
(\delta_{ia}\delta_{jb}+\delta_{ja}\delta_{ib})\end{equation}
The contribution of $\beta$ to the mass of the gauge bosons reads (with
wave function renormalization $Z_\beta$)
\begin{equation}\label{QA}
M_\rho^2=2Z_\beta g^2\bar\beta^2,\quad M^2_{K^*}=Z_\beta g^2\bar
\beta^2,\quad M^2_\omega=\frac{2}{3}Z_\beta g^2\bar\beta^2\end{equation}
All local symmetries are spontaneously broken and all gauge bosons
are massive. As for the case of the $\delta$-condensate, the
$\beta$-condensate leaves the global symmetry associated to $B+S$
unbroken. Again, a $\beta$-condensate does not lead to a massless
Goldstone boson and to superfluidity. Actually, the fields
$\bar\delta$ and $\bar\beta$ (\ref{I}), (\ref{Q}) carry the same abelian
quantum numbers and are both isospin singlets. As before, a gauged electromagnetic symmetry
remains preserved by a $\beta$-condensate. In presence of an octet
condensate the antitriplet
and sextet can mix. More precisely a color octet and antitriplet
diquark condensate induce a sextet diquark condensate
by a linear term in the effective potential for the sextet.
This term is generated by a cubic interaction between
$\chi,\delta$ and $\beta$. Similarly, $<\chi>\not=0$ and $<\beta>\not=0$
induce $<\delta>\not=0$.

In summary, the Higgs picture for two-flavor QCD differs in
two important aspects from the three-flavor case. First, the vacuum either
admits diquark condensation or one finds
a massless gauge boson in the vacuum (in addition to the photon).
This holds if electric charges are integer and isospin is conserved.
Second, there is no massless Goldstone boson in presence of
diquark condensates. Therefore no superfluidity occurs in the high
density phase if isospin is conserved.

\section{Gluonic condensates}

We conclude the discussion of isospin conserving
states in QCD$_2$ by a brief listing of
possible color-breaking gluon condensates. As long as
they conserve $SU(2)_I$ and $B+S$, they are necessarily
induced in presence of color breaking quark-antiquark and diquark
condensates. The simplest $SU(2)_I$-singlet operator is
found in the octet
\begin{equation}\label{X1}
f_{ij}\sim(F^{\mu\nu}_{ik}F_{kj,\mu\nu}-\frac{1}{3}
F^{\mu\nu}_{lk}F_{kl,\mu\nu}\delta_{ij})\end{equation}
With respect to $SU(2)_I$ it decomposes as $8\to 1+2+2+3$, with
a possible singlet expectation value
\begin{equation}\label{X2}
<f_{ij}>=\bar f(\lambda_8)_{ij}\end{equation}
This remains neutral with respect to $U(1)_8$ and therefore cannot
break this local symmetry. Nevertheless, it gives a contribution to
$M_{K^*}$, i.e. it contributes
\begin{equation}\label{X3}
M^2_\rho=0\ ,\ M^2_\omega=0\ ,\
M^2_{K^*}=\frac{3}{2}Z_fg^2\bar f^2 \end{equation}
The other nontrivial candidate in the (symmetric) product of two-field
strength tensors is the 27-dimensional representation
\begin{eqnarray}\label{X4}
&&s_{ijkl}\sim F^{\mu\nu}_{ij}F_{kl,\mu\nu}-\frac{1}{3}(F^{\mu\nu}_{im}
F_{ml,\mu\nu}\delta_{jk}\nonumber\\
&&\qquad+F^{\mu\nu}_{km}F_{mj,\mu\nu}\delta_{il}
-\frac{1}{3}F^{\mu\nu}_{mn}F_{nm,\mu\nu}
\delta_{ij}\delta_{kl})\nonumber\\
&&s_{ijjl}=s_{ijki}=0\end{eqnarray}
It contains one $SU(2)$-singlet which is again invariant
with respect to $U(1)_8$ and contributes to the gluon masses
similar to (\ref{X3}).

The lowest dimension $SU(3)_c$ representation which contains
a $SU(2)_c$ singlet with nonzero $U(1)_8$-charge and which has
triality zero is the complex 10. It is contained in the antisymmetric
product of two octets. A scalar formed from glue in the $10+\bar{10}$
respresentation is
\begin{eqnarray}\label{X5}
a_{ijkl}&\sim&F^{\mu\nu}_{im}F_{mj,\nu}^{\ \ \ \ \rho}
F_{kl,\rho\mu}-
F^{\mu\nu}_{km}F_{ml,\nu}^{\ \ \ \ \rho}
F_{ij,\rho\mu}\nonumber\\
&&-\frac{1}{3}F^{\mu\nu}_{nm}F_{mn,\nu}^{\ \ \ \ \rho}
F_{kl,\rho\mu}\delta_{ij}
+\frac{1}{3}F^{\mu\nu}_{nm}F_{mn,\nu}^{\ \  \ \ \rho}
F_{ij,\rho\mu}\delta_{kl}\end{eqnarray}
with $a_{ijkl}=-a_{klij},\ a_{ijjl}=a_{ljkl}
=0$.
An expectation value of the $SU(2)$-singlets in $a_{ijkl}$ would break
$U(1)_8$ and give a mass to the corresponding gauge boson. A condensate
of such a higher order gluonic operator would also break $B+S$. Only
$SU(2)_I$ and baryon number remain as unbroken global symmetries
of the vacuum if such operators acquire an expectation value. After
a coupling to electromagnetism the baryons would get the noninteger
charges corresponding to $Q'=\tilde Q-\frac{1}{2}\lambda_3$
(see table 2). In presence
of such a $B+S$-violating condensate the occurrence of a diquark
condensate in the high density phase would break the only global
symmetry, i.e. baryon number. ($B+S$ is not ``available'' for a
residual global symmetry any more). The resulting Goldstone boson
would lead to superfluidity. For completeness we have listed the characteristics of
$a$-condensates in table 1 (entries F, G). We will not consider them
any further since fractional electric charges in the vacuum seem not
very attractive.

\section{Transition between QCD$_3$ and QCD$_2$}

The vacuum of QCD with three light flavors of quarks exhibits
the vectorlike $SU(3)$-symmetry of the ``eightfold way'', if the
quark masses are all equal. No diquarks can condense since this
would break either baryon number or the $SU(3)$ symmetry. The only
nontrivial condensate is a color octet quark-antiquark pair
\cite{GM}. The ground state of realistic QCD presumably resembles
three-flavor QCD and does not exhibit a diquark condensate either.
This follows from the fact that the global symmetries corresponding
to baryon number and strangeness are realized in a standard way.
Both $B$ and $S$ are conserved quantum numbers, in contrast to the
case of diquark condensation.

For a small mass difference between the strange quark and the two
light quarks we expect a splitting of the masses of particles
within a given $SU(3)$-multiplet according to their isospin
representation. The color singlet and octet $\bar qq$-condensates
involving strange quarks add to eq. (\ref{C}) a
contribution\footnote{We have omitted here one more electrically
neutral $SU(2)_I$-singlet contained in the color octet, $\xi_5$.
The expectation value $\bar\xi_5$
vanishes in the limit of equal quark masses.}
\begin{eqnarray}\label{S1}
\Delta\sigma_{ab}&=&\bar\sigma_s\delta_{a3}\delta_{b3}\nonumber\\
\Delta\xi_{ij,ab}&=&\frac{1}{2\sqrt6}\bar\xi_3(\lambda^4_{ij}\lambda^4_
{ba}+\lambda^5_{ij}\lambda^5_{ba}+\lambda^6_{ij}
\lambda^6_{ba}+\lambda^7_{ij}\lambda^7_{ba})\nonumber\\
&&-\frac{1}{3\sqrt2}\bar\xi_4\lambda^8_{ij}\delta_{a3}\delta_{b3}
\end{eqnarray}
The additional condensates $\bar\sigma_s,\bar\xi_3,\bar\xi_4$ involve
the third flavor index\footnote{Remember that in eq. (\ref{C}) the flavor
index takes only the values 1,2 whereas now it runs from 1 to 3.}
and will be denoted as ``strange components'' of the color singlet
and octet. Combining the ``strange'' and ``nonstrange'' octet
contributions to the vector meson masses one obtains
\begin{eqnarray}\label{S2}
M_\rho^2&=&\hat Zg^2(\frac{2}{3}|\bar\xi_1|^2+\frac{1}{3}|
\bar\xi_3|^2)\nonumber\\
M^2_{K^*}&=&\hat Z g^2(\frac{1}{4}|\bar\xi_1|^2+\frac{1}{12}|\bar\xi_2|^2+\frac{1}{2}
|\bar\xi_3|^2+\frac{1}{6}|\bar\xi_4|^2)+\frac{3}{2} Z_fg^2\bar f^2
\nonumber\\
M^2_\omega&=&\hat Z g^2|\xi_3|^2\end{eqnarray}
In the $SU(3)$-symmetric limit $\bar\xi_1=\bar\xi_2=\bar\xi_3=\bar\xi_4
,\bar f=0$
the masses in the vector meson octet are all degenerate \cite{GM}.
We observe that for $\bar\xi_1=\bar\xi_3$ the masses of the $\rho$-mesons
and $\omega$-meson remain equal.
For $m_s\not= m_{u,d}$ one expects that also colored gluon
condensates of the type (\ref{X2}) are induced. If the octet
condensate $\sim\bar\xi_3-\bar\xi_1$
is suppressed as compared to
the other $SU(3)$-breaking condensates,
we find the phenomenologically interesting relation
$M^2_\omega\approx M_\rho^2, M_{K^*}^2>M^2_\rho$.

It is, of course, possible
to move gradually from three-flavor QCD to two-flavor QCD by
increasing the mass of the strange quark. In this process the
various $SU(2)_I$-singlets acquire
different expectation values. For a small deviation from $SU(3)$ this leads
to a mass splitting in the $SU(3)$-multiplets according to
strangeness. In the limit $m_s\to\infty$ the condensates
involving strange quarks vanish. In this limit the $\omega$-meson
would become massless in the (naive) Higgs picture for any vacuum
without a diquark condensate.
In this case it is conceivable, but not
likely that a phase transition occurs as a function of $m_s$
where the local $U(1)_8$-symmetry gets restored for large enough $m_s$ and
the $\omega$-meson becomes massless. At this point the reader should be warned,
however, that the naive Higgs picture with bosonic propagator approximated
by $(q^2+m^2)^{-1}$ could be quite misleading in presence of strong couplings.
It is well possible that the propagator for small momenta and $m^2=0$ looks very
different from $1/q^2-$ e. g. $1/q^4-$ such that the analytic continuation to
Minkowski space has no pole. In this case no propagating gluon-type degree of freedom
would occur and there would be no massless $\omega$-meson. In this sense the ``naive''
Higgs picture underlying our present discussion may only be valid as long as the
effective bosonic mass $m^2$ is large enough. In view of the fact that couplings are
strong and no global symmetries are altered for $m_s\rightarrow\infty$ we find a
phase transition not very likely in absence of a diquark condensate in the QCD$_2$ vacuum.

From the point of view of a possible phase transition the
perhaps more interesting alternative is characterized
by a vacuum diquark condensate (\ref{I}) in the two-flavor limit
$m_s\to\infty$. In this event two scenarios imply
a phase transition between
QCD$_3$ and QCD$_2$ as a function of $m_s$.
Either the diquark condensation sets in in competition to the ``strange'' octet components.
At the transition the conserved quantum numbers jump from $B$ and $S$ to the
new charges $B'$ and $S'$. Or their could be an intermediate
phase with both diquarks and all components of the $q\bar q$-octet
condensing. This intermediate phase would be signalled by
superfluidity connected to the Goldstone boson arising from a
reduction of the total number of global symmetries. In the latter case a phase
transition would be mandatory since the global symmetries change as a function of $m_s$.

Keeping in mind our cautious remarks we next investigate the transition between large
and small $m_s$ within the naive Higgs picture. From the point of view
of QCD$_2$ we may first add a heavy strange quark to QCD$_2$. This will not
affect the low energy dynamics and the condensates of QCD$_2$. It
introduces, however, a new global symmetry with conserved
quantum number $S'$
for the heavy quarks, i.e. $S'=-1$ for all strange quarks
and $S'=0$ for up and down quarks. The quantum numbers of the $s$-quark
states in the hadronic language can be found in table 4, which
supplements table 2.

\medskip
\begin{center}
\begin{tabular}{|c|ccc|ccc|cc|c|}
\hline
&$\tilde Q$&$Q_c$&Q&$I_3$&$S$&$B$&$B'$&$S'$&\\
\hline
$s_1$&-1/3&2/3&-1&-1/2&-2&1&-1&-1&$\Xi^-$\\
$s_2$&-1/3&-1/3&0&1/2&-2&1&-1&-1&$\Xi^0$\\
$s_3$&-1/3&-1/3&0&0&-1&1&0&-1&$S^0$\\
\hline
\end{tabular}\\

\medskip Table 4:  Charges of the strange quarks
\end{center}

We start with the case where the vacuum of QCD$_2$ exhibits
a nonvanishing diquark condensate.  In presence
of this diquark condensate  the conserved
quantum numbers for large $m_s$ are $Q, B'$ and $S'$. For
small $m_s$, the vacuum must switch, however, to a state which
conserves $Q, B$ and $S$, similar to QCD$_3$. As we have mentioned, such
a state is not compatible with a diquark condensate. The phase
transition as a function of $m_s$ corresponds
therefore to the disappearance of the diquark condensate
and to an associated change in
quantum numbers of states in the low energy spectrum. It is supposed
to happen for some  critical
value $m_{s,c}$. This change in the quantum numbers
concerns all ``strange baryons'',
i.e. not only the ones corresponding to strange quarks (table 4),
but also four of the ``light quark states'' listed in table 2. It
is noteworthy
that the proton and neutron or the $\rho$-mesons
are not affected by this switch in quantum numbers.

The switch in quantum numbers from $B'$ and $S'$ to $B$ and $S$ may
either proceed directly or via an intermediate superfluid or
Coulomb-like phase. The reason is the mismatch in quantum numbers
for the diquark condensate and the ``strange'' component of the
octet $\bar qq$-condensate. A transition where a nonzero diquark condensate
is replaced by nonzero strangeness components of the octet condensate
at the critical strange quark mass leads to a jump in the conserved
charges. If the $\omega$-meson is massless for $m_{s,c}$ or even
for a whole interval of the strange quark mass, no discontinuity is
necessary. In the intermediate Coulomb phase the $U(1)_8$ local gauge
supersymmetry associated to the $\omega$-meson allows one to rotate
freely between $(B,S)$ and $(B',S')$. Both pairs of quantum numbers are
conserved. The limits of large and small $m_s$ outside this
``Coulomb''-region correspond to different lockings of the $\lambda_8$
generator, whereas in the Coulomb region it is unlocked. (Recall that
beyond the naive Higgs picture the ``Coulomb region'' may not have a
massless gluonic excitation.) If the change from
the diquark condensate to the condensation of the strange components of the
octet proceeds without an intermediate ``Coulomb region'', it may be
most likely a discontinuous first-order phase transition. Finally,
if there is an intermediate range of $m_s$ where both the diquark
condensate and the strange component of the octet condensate are nonzero,
there simultaneous presence leads to the breaking of one of the two
global symmetries. This corresponds to an intermediate superfluid
phase.

Finally, we discuss the perhaps less likely possibility of a phase transition
without diquark condensation in the vacuum
of QCD$_2$. In this case one starts at high $m_s$ with a Coulomb-like phase
where both $(B',S')$ and $(B,S)$ are
simultaneously conserved. The transition at $m_{s,c}$ is then
characterized by the onset of
a nonvanishing vacuum expectation value of the strange components
of the octet. This gives a mass to the $\omega$-meson and
the transition shows similarities to the transition in
a superconductor.

All versions of the transition show a remarkable behavior of the
fate of the strange baryons as $m_s$ increases from zero to
large values. Only the $\Xi$-baryons and the singlet $S^0$
become very heavy and decouple from the effective low
energy theory, similar to the baryons involving the heavy quarks
$c, b$ and $t$ \cite{GM}. This decoupling does not take
place for the baryons $\Sigma$ and $\Lambda^0$ and for the $K^*$-vector
mesons. These particles remain in the spectrum of light particles
even in the limit $m_s\to\infty$. If diquark condensation sets
in for large $m_s$, their properties become unusual since they
cannot be build any more from a finite number of quarks in the
nonrelativistic quark model \cite{BerW}.

This unusual behavior of the $K^*,\Sigma$ and $\Lambda$ particles
may give a hint where to look for a possible transition in a lattice
simulation. Indeed, the masses and couplings of the pseudoscalars
$(\pi, K,\eta,\eta^{\prime})$, the $\rho$-vector mesons and the nucleons $(p,n)$ may
depend rather smoothly on $m_s$ (with the possibility of minor jumps in case of
a discontinuous transition). In particular, we have seen that the masses of $p,n,\rho,\pi$
are not affected by a diquark condensate. The qualitative changes between $m_s<m_{s,c}$ and
$m_s>m_{s,c}$ are rather expected in the $(K^*, \Sigma,\Lambda)$ sector.
Here one may find some unusual behavior as the mass difference $m_s-m_q$ increases
from (realistic) values within the approximate validity of the SU(3) flavor symmetry
to large values. (Note that the relevant parameter is actually $m_s-m_q$ with $m_q$
the mass of the up/down quark. It is important to keep this in mind since lattice
simulations with very small $m_q$ are not feasible at the present moment.)
The present simulations with three dynamical quarks \cite{LSDQ} are presumably within the
range of approximate $SU(3)$ symmetry. It would be interesting to increase $m_s$ with
fixed $m_q$ and to monitor the behavior of $(K^*,\Sigma,\Lambda)$.

We observe,
however, that for $m_s<m_{s,c}$ the ``dressed gluons'' $K^*$ can mix with the
corresponding quark-antiquark vector states $\sim\bar{q}\gamma^{\mu}q$ and similar for the
``dressed quarks'' $\Sigma,\Lambda$ which can mix with three quark states. \footnote{
See ref. \cite{GM} for a detailed discussion of the quantum numbers of the dressed
gauge invariant states.}
This mixing may be forbidden for $m_s>m_{s,c}$, as in the case of diquark condensation.
The correlation functions for the usual $\bar{q}\gamma^{\mu}q$ or $qqq$ operators may
therefore not reflect anymore the states that we have denoted by
$(K^*,\Sigma,\Lambda)$ in the Higgs picture. This feature may render a direct
observation of a transition rather difficult as long as one concentrates on correlation
functions of operators like $\bar{q}\gamma^{\mu}q$ and $qqq$. In case of diquark
condensation in the vacuum a more direct access to the particular features of such a
ground state would become possible if local gauge invariant operators corresponding to the
states $K^*,\Sigma,\Lambda$ can be constructed. These must be nonlinear in the
quark and gluon fields since no mesonic state with half integer isopin can be constructed
as a finite power of quark and gluon fields. This also holds for $\Sigma$ and $\Lambda$ which
have even isospin and should only involve contributions with an odd number of fermions.

\section{High-density phase transition in QCD$_2$}

\medskip
The possible phase transitions between the vacuum and the high
density state of QCD$_2$ depend on the condensates in the vacuum
and  the high density phase. We summarize in table 5 the possibilities
for the case of an isospin-conserving high density phase with octet
$\bar qq$- and diquark condensate. The global
symmetries of the high-density state of
QCD$_2$ and the vacuum are the same. In this event a continuous
transition becomes possible. We distinguish the two alternatives
without and with a diquark condensate in the vacuum ((A) and (B) in
table 1). For the latter case there is no obvious reason for a phase
transition. On the other hand, if the transition to the high density
phase is characterized by the onset of diquark condensation
one may expect a transition similar to the abelian Higgs
model. This could be a first-order transition, but a second-order
transition or a continuous crossover are also conceivable \cite{AbH}. For
a weak enough transition one expects the same universality class as for type
II superconductors.

\medskip
\begin{center}
\begin{tabular}{|c|c|c|c|c|}
\hline
vacuum&high&``phase''&high&massless\\
&density&transition&density&gluon in\\
&&&superfluidity&vacuum\\
\hline
$8$&$8+\bar 3,6$&abelian&no&yes\\
&& Higgs&&\\
$8+\bar 3,6$&$8+\bar3,6$&continuous\ or&no&no\\
&&first order&&\\
\hline
\end{tabular}\\

\medskip Table 5: Possible isospin conserving high density transition  in
QCD$_2$
\end{center}

One may argue, however, that these scenarios are too simple.
It seems likely that effective attractive interactions
between the nucleons lead to an instability of the
isospin conserving states discussed so far. This
instability is related to the pairing of nucleons.
There is no contribution to a Majorana like mass term
for the proton and neutron from the diquark condensates $\bar\delta$
(cf. eq. (\ref{16B}) or $\bar\beta$.)
In fact, $s$-wave and spin 0 nucleon pairs belong to an isospin triplet.
This follows from the Pauli-principle since the color part of
the corresponding diquark operator is symmetric (proton and
neutron are both quarks of the third color, see table 2) and
the spin part is antisymmetric. The dinucleon operator must
therefore be symmetric in flavor, and this corresponds to an
isospin triplet. Since the isospin singlet diquarks $\bar\delta$ and
$\bar\beta$ cannot
stabilize an instability in the nucleon pair channel, it seems
likely that the condensation of nucleon pairs or diquarks
of the third color produces the stabilizing gap.

We suggest that isospin is spontaneously broken in the high
density phase of QCD$_2$
as well as in nuclear matter.
A candidate is a condensation of a diquark
\begin{equation}\label{D1}
<d_3c d_3>\not=0\end{equation}
In the language of baryons this corresponds to a dineutron
condensate. Such a condensate has interesting consequences. First,
the spontaneous breaking of isospin produces three massless Goldstone
bosons. In contrast to the pions they are scalars with quantum numbers
of the $a$-mesons. In presence of a nonvanishing
mass difference for the (current) mass of the up and down quark (or
in presence of
electromagnetism) the global isospin symmetry is explicitly
broken. As a consequence, the two charged $a^\pm$ mesons become
pseudo-Goldstone bosons and acquire a small mass, similar to the pions.
In contrast, the third component of isospin $I_3$ remains an exact
symmetry even in presence of quark masses and electromagnetism.
Its spontaneous breaking by the dineutron condensate
(\ref{D1}) necessarily
leads to an exactly massless Goldstone boson and therefore to
superfluidity. Second, also the baryon number $B'$ is not conserved
by the dineutron condensate. The only unbroken generator is
$I_0+\frac{1}{2}B'$ and corresponds to conserved electric
charge $Q$.
The simultaneous presence of three light pseudoscalars  $\pi$
and three light scalars $a$ is a characteristic signal for
our scenario with spontaneous breaking of the chiral and vector-like
global $SU(2)$-symmetries at high density.

Protons and neutrons are presumably the lightest fermionic
states in the vacuum of QCD$_2$. As the chemical potential for
the conserved baryon number increases beyond a critical
value the onset of the dineutron condensate triggers a true
phase transition. The order parameter is related to the
spontaneous breaking of the global symmetry with generator $I_3$.
This is the ``gas-liquid'' transition to nuclear matter. If this
phase transition would be of second order, it should belong to a
universality class characterized by the breaking of $SU(2)$ to
$U(1)$  (similar to $O(3)$-Heisenberg models) with small
symmetry violations due to $m_u\not=m_d$. The fermionic
fluctuations presumably play a role for this universality
class. It is an open issue if for still higher values of the
chemical potential another phase transition to high density
quark matter is connected to the changes in the isospin singlet
diquarks shown in table 5 or if the transition from nuclear to
quark matter is continuous.

In the following we summarize the main features of the high
density phase transition for the case where spontaneous
isospin symmetry breaking is absent or can be treated as a
small correction.

(i) At high density complete color flavor
locking is possible such that all gluons acquire a mass
\cite{BerHD}. This can be realized
by the same combination of two condensates as in the vacuum.
If isospin is preserved there is no Goldstone boson and the high
density phase is not a superfluid. Chiral symmetry remains
spontaneously broken at high density.
Without spontaneous isospin breaking the transition between the
vacuum and the high density state of QCD$_2$ would not involve
any change in the realization of symmetries
if diquarks condense already in the vacuum. This leaves the possibilities
of an analytic  crossover or a first-order phase transition.

(ii) In an alternative scenario for the QCD vacuum the diquark
condensate could vanish. In this case
not all gluons can acquire a mass. There
remains always an unbroken abelian $U(1)_c$ gauge
symmetry which is part of $SU(3)_c$. Its gauge boson carries the quantum
numbers of the $\omega$-meson and remains massless.
If, in the second scenario,  the transition to QCD at
high baryon density corresponds to condensation of a
diquark which preserves isospin, one infers that
the $U(1)_c$-gauge symmetry
gets  spontaneously broken in the transition to the
high density phase. The $\omega$-meson acquires a mass.
If isospin is conserved, the general characteristics
of the high density
phase transition for the second scenario may resemble the transition
for the abelian Higgs model \cite{AbH}. This may
lead to a first-order phase transition. The remarks on the limitations of the
naive Higgs model from the preceeding section apply here as well. In particular, it is
conceivable that no discontinuity occurs and the vacuum is analytically connected
to the high density state. The number of global abelian
symmetries is the same with and
without isospin singlet diquark condensates. Therefore no Goldstone boson is generated
by an isospin-conserving diquark condensation in a transition
to the high density phase.

\section{Gluodynamics}

Finally, we turn to pure QCD without quarks. A description in the
Higgs picture with spontaneous breaking of color may still be
possible. Of course, it is not related to color-flavor locking
any more, since there are no flavor symmetries.
We present here a first attempt which demonstrates
that suitable condensates can give a mass to all gluons.
The resulting spectrum for the glueballs
is, however, not very satisfactory. Our example
should therefore not be interpreted as a proposal
for the ground state of pure QCD but rather as an exploration
in which directions one might go. For our first trial we follow the same
philosophy as for the fermion bilinears and introduce scalar fields
$f_{ij}$ and $s_{ijkl}$ for the color nonsinglets
contained in $F^{\mu\nu}F_{\mu\nu}$ (cf. eqs. (\ref{X1}), (\ref{X4})).
We assume that the effective action for these scalar fields
\begin{equation}\label{G0}
{\cal L}=\frac{1}{2} Z_f(D^\mu f)_{ij}(D_\mu f)_{ji}+\frac{1}{2}
Z_s(D^\mu s)_{ijkl}(D_\mu s)_{jilk}+U(f,s)\end{equation}
has a potential $U$ with minimum for nonzero expectation values
of $f$ and $s$ (see \cite{Reu} for a similar treatment of the color
singlet). We want to demonstrate here that for suitable expectation
values all gluons become massive and can be associated with vector
glueballs.

We first consider the color octet $f_{ij}=(f^\dagger)_{ij}, f_{ii}=0$.
By appropriate $SU(3)$-transformations its vacuum expectation
can be brought to the generic form
\begin{equation}\label{G1}
<f_{ij}>=\bar f(\lambda_8)_{ij}+\bar t(\lambda_3)_{ij}\end{equation}
Such an expectation value cannot give mass to all gluons -- the $A_3$-
and $A_8$-vector mesons remain massless. For $\bar t\not=0$ isospin symmetry
is not conserved and the vector meson masses split according to
\begin{eqnarray}\label{G2}
&&M^2_1=M^2_2=2Z_f g^2\bar t^2\quad,\quad M^2_3=0\quad,\nonumber\\
&&M^2_4=M^2_5=M^2_6=M^2_7=\frac{1}{2}Z_fg^2(3\bar f^2+\bar t^2)\quad,
\quad M^2_8=0\end{eqnarray}

Giving a mass to all gluons therefore needs one more nonvanishing
expectation value and we consider here a particular direction
in the 27-dimensional representation $s_{ijkl}=s^*_{jilk}$, namely
\begin{equation}\label{G3}
<s_{ijkl}>=\bar s(\delta_{ik}\delta_{jl}-\frac{1}{3}\delta_{il}
\delta_{jk})\end{equation}
This adds a mass term to the gluons corresponding to the symmetric
Gell-Mann matrices
\begin{eqnarray}\label{G4}
&&M^2_1=M^2_3=M^2_4=M^2_6=M^2_8=12Z_s g^2\bar s^2\nonumber\\
&&M^2_2=M^2_5=M^2_7=0\end{eqnarray}
Combining (\ref{G4}) with (\ref{G2}) all gluons have acquired
a mass. We also observe the split between $M^2_1$ and $M^2_2$ etc.

We would like to associate the massive gluons with vector glueballs.
Together
with the scalar glueballs described by $f, s$ and a
corresponding color singlet they would be
expected to dominate the low
energy spectrum of pure QCD if the vacuum can be characterized
by eqs. (\ref{G1}) and (\ref{G3}). (Spin 2 glueballs can be
described by introducing additional fields for operators
like $F_\mu^{\ \rho}F_{\rho\nu}$.) In order
to differentiate between the discrete transformation properties
of the glueballs we need the action of the parity transformation
and charge conjugation
\begin{equation}\label{G5}
P:\quad A_0\to A_0,\quad A_i\to-A_i,\quad C:\quad A_\mu
\to -A^T_\mu\end{equation}
where $P$ is accompanied by a coordinate reflection. The ``symmetric gluons''
$A_1,A_3,A_4,A_6,A_8$ transform as $1^{--}$ vector glueballs,
whereas the ``antisymmetric gluons''
$A_2,A_5,A_7$ correspond to $1^{-+}$ vector glueballs.

For the scalars we exploit
\begin{eqnarray}\label{G6}
P: && F^{ij}\to F^{ij}\ , \quad F^{0j}\to -F^{0j}\ ,\quad f_{ij}\to
f_{ij}\ ,\quad s_{ijkl}\to s_{ijkl}\nonumber\\
C:&& F^{\mu\nu}\to -(F^T)^{\mu\nu}\ ,\quad f_{ij}\to f_{ji}\ ,\quad
s_{ijkl}\to s_{jilk}\end{eqnarray}
This shows that the expectation values (\ref{G1}), (\ref{G3})
indeed conserve $P$ and $C$. The scalar glueballs consist of
$0^{++}$ and $0^{+-}$ states. The detailed mass spectrum of the
vector and scalar glueballs depends on the properties of the
effective potential $U(f,s)$ and may involve expectation values beyond
those considered in (\ref{G1}), (\ref{G3}). The effective
action (\ref{G0}) with condensates (\ref{G1}) and (\ref{G3})
nevertheless predicts a selection rule, namely the absence
of $1^{++}, 1^{+-}, 0^{-+}$ and $0^{--}$ states
among the lightest glueball states.
This has to be confronted with the results of lattice simulations
\cite{Teper} which find the lightest glueball states (with
increasing mass) as $0^{++}, 0^{-+}, 2^{++}, 1^{+-}$. The lack
of agreement shows that our first attempt has not been successful.
Since the failure is linked to the action of the discrete
symmetries $P$ and $C$ and the rotation group on the
gauge fields $A_\mu$, it concerns all other possible condensates
which preserve these symmetries as well. An interesting way out of this
dilemma may combine rotations with gauge symmetries or similar for the
discrete symmetries, corresponding to a type of
``color-Lorentz'' or ``color-spin'' locking \cite{QCDGS}.\footnote{See
\cite{CSL} for a discussion of color-spin locking for one-flavor
QCD at high density.}

Of course, it is not necessary that gluodynamics admits a Higgs description.
It is well conceivable that the gluon propagator cannot be characterized by a mass term
generated from ``spontaneous symmetry breaking''. This does not mean that the gluons will
manifest themselves as massless excitations. The gluon propagator may simply admit
no particle pole. \footnote{These statements are meaningful only for
some appropriate gauge fixing. On the other hand, it is not easy to see why a particle
pole in a fixed gauge would not correspond to a physical particle in a gauge invariant
setting.} In any case, we expect that the mechanism which removes the perturbatively
massless gluons from the very low momentum spectrum is very different for gluodynamics
and for realistic QCD$_3$. We therefore
emphasize that the role of the gluons and glueballs in realistic QCD$_3$ with
three light quark flavors is quite distinct from gluodynamics. In our Higgs picture of
realistic QCD the gluons are associated with the $(\rho,K^*,\omega)$-mesons
and do not correspond to glueball states.
Glueballs play an important role in
the low energy spectrum of gluodynamics, whereas for QCD$_3$ they may
only appear as some higher excited states. In our picture the origin of this
different role of the gluons and glueballs are the different patterns in the spontaneous
breaking of the color symmetry.

\section{Transition to gluodynamics and heavy quark potential}

Our discussion of gluodynamics in the previous section raises the question
what happens if the mass off all three light quarks is increased simultaneously.
For simplicity we consider an equal mass $m$ for the up, down and strange quarks.
For large enough $m$ the low momentum sector is characterized by gluodynamics. We therefore
expect a transition from realistic QCD$_3$ to gluodynamics as $m$ increases from small to
large values. As we have discussed in the previous section this should be associated with
a change of the effective infrared cutoff. Indeed, one may imagine that the infrared cutoff
for QCD$_3$ provided by the octet induced gluon mass competes with some other, not yet well
identified cutoff which is relevant for gluodynamics. We propose that
for small enough $m$ the octet
condensate sets the largest IR-scale and the effective cutoff for gluodynamics is therefore
ineffective. For example, a possible nontrivial momentum dependence of the gluon propagator
for ``massless'' gluons may not get realized since the fluctuations responsible for it are cut
off by the gluon mass term. On the other hand, for $m$ above some critical value $m_{cr}$
the IR-cutoff of gluodynamics dominates. In turn, it may cut off the fluctuations that
could be responsible for the octet condensate as, for example, the large size instantons
\cite{IN}. For very large $m$ the color octet $\bar{q}q$-bilinear simply
plays no important role.
The ``switch'' between the relevant cutoffs in some mass region around
$m_{cr}$ may be a sharp or smooth
crossover since no global symmetries are affected. It would be interesting to have at least
a rough idea about the value of $m_{cr}$. It seems reasonable that pion and kaon masses
below $200-300MeV$ are already close to the chiral limit and therefore correspond to
$m<m_{cr}$. On the other hand, for both pseudoscalar and vector masses above $1GeV$ one is
presumably in the ``heavy quark region'' $m>m_{cr}$.

Lattice simulations \cite{LSDQ} show a rather smooth behavior of the spectrum if the
pseudoscalar masses are varied between $300MeV$ and $1~GeV$. This is consistent with a
smooth crossover and seems to exclude any phase transition between QCD$_3$ and gluodynamics.
We find it not unlikely that the crossover region where the effects of octet condensation
start to become important corresponds to pseudoscalar masses in the vicinity of $500MeV$.
As a consequence, this mass region would not yet belong to the range where chiral
perturbation theory applies. One may wonder where to expect the most prominent signs of the
octet condensation once $m$ decreases below $m_{cr}$. One is perhaps the influence of the
octet on the value of $f_{\pi}$. Another one concerns the quark mass dependence of the vector
meson masses. The first involves the details of the interplay between the octet
and singlet $\bar{q}q$ contribution as well as the quark mass dependence of the
pseudo-scalar wave function renormalization. The second needs an understanding of the
mixing between gluons and $\bar{q}\gamma^{\mu}q$.

Another interesting issue is the shape of the heavy quark potential
(for $c$ or $b$ quarks). Indeed, for $m$ sufficiently large we expect
a linearly rising potential $V(r)$ in a certain region of $r$, while for very large $r$
the potential flattens due to string breaking. The slope of the linear rise is associated
to the string tension. On the other hand, there may be a region of small $m$ for which the
linearly rising piece in the potential is absent. It is an interesting hypothesis that the
potential could be described in this region by a naive Higgs picture with a
Yukawa type potential
\begin{equation} \label{8.1}
\frac{\partial V(r)}{\partial r}=\frac{4\hat{\alpha}_s(r)}{3r^2}\exp(-\bar{M}_{\rho}r)
\end{equation}
Here $\bar{M}_{\rho}$ is the gluon mass associated to an average mass of the vector mesons
$(\rho,K^*,\omega)$ and $\hat{\alpha}_s(r)$ corresponds to some suitably defined
strong fine structure constant. The value of $\bar{M}_{\rho}$ depends on $m$ and the precise
shape of $\alpha_s(r)$ is influenced by $\bar{M}_{\rho}$ for the region
$r\geq \bar{M}_{\rho}^{-1}$.

We point out that the string picture and the Yukawa potential are not incompatible - this
reflects the duality between a confinement and a Higgs description.
For a small enough quark mass $m$ it is cconceivable
that $V(r)$ can both be described by the Yukawa potential
(\ref{8.1}) with reasonable $\hat{\alpha}_s(r)$ and by a string breaking picture. On
the other hand, for large $m$ the Yukawa description is expected to break down
and only the string picture remains valid. In the limit $m\rightarrow\infty$ one
then recovers gluodynamics without string breaking.

Let us first discuss the qualitative behavior of $V(r)$ for very large $r$ in the string
picture. For finite $m$ there will be a ``string breaking scale'' $r^{-1}_B$ such that
the potential approaches quickly a constant for $r>r_B$. The approach to the constant is
expected to be exponential
\begin{equation} \label{8.2}
\frac{\partial V}{\partial r}=\sum_ic_i(r)\exp(-M_ir)
\end{equation}
For large enough $r$ eq. (\ref{8.2}) will be dominated by the masses $M_i$ of the lightest
mesons which are exchanged between heavy quarks (or heavy charmed or beauty baryons or
mesons). Neglecting the pions and other pseudoscalar mesons (whose contribution should
also be added to eq. (\ref{8.1})) this is precisely the behavior of eq. (\ref{8.1})
with $M_i=\bar{M}_{\rho}$. For very large $r$ beyond the string breaking distance $r_B$
we expect always a Yukawa type force, independent of the precise mechanism generating
the meson masses. This region in the potential therefore finds a similar description
in the Higgs picture and the string breaking picture.

A Yukawa type description is excluded, however, if the string breaking distance $r_B$ is
large enough, say $r^{-1}_B{>\atop\sim}300MeV$ or $r_B{>\atop\sim}0.7$ fm.
Here we define $r_B$ as the upper end of a range $r<r_B$ for
which $\partial V/\partial r\approx \sigma$, with $\sigma$ the constant string tension.
In fact, for $\bar{M}_{\rho}r_B\gg 1$ one concludes that there exists a range in $r$ where the
potential is linearly rising and simultaneously $\bar{M}_{\rho}r\gg 1$. This is in
contradiction to the exponential suppression for the ``pure Yukawa potential''
(\ref{8.1}). On the other hand, for $r^{-1}_B\rightarrow ~400MeV$ the linear piece in the
potential disappears. In this range of $r$ the exponential suppression due to
the mass is not yet dominant and the naive Higgs picture with Yukawa potential (\ref{8.1})
could become valid.

For large enough $m$ the string breaking distance $r_B$ can be roughly
estimated from a simple
energy argument. The potential energy for the separation of two heavy quarks with mass
$m_H$ is $V=\sigma r$. The string will break once $2m_H+\sigma r$ is large enough in order
to produce twice the mass $m_B$ of a heavy-light meson (e. g. D or B meson). This yields
\begin{equation}
r_B\approx\frac{2(m_B-m_H)}{\sigma}
\end{equation}
For very large $m$ one has $m_B\approx m_H+m$ and therefore
$\bar{M}_{\rho}r_B=2m\bar{M}_{\rho}/\sigma$. Simplifying further $\bar{M}_{\rho}\approx 2m$
yields as a condition for $m>m_{cr}$ the inequality $2m\gg\sqrt{\sigma}$ or $m\gg200~MeV$.
This simple estimate may become valid for $m{>\atop\sim}500MeV$. Present lattice simulations
with pion masses above $300-400 MeV$ have seen no sign of string breaking yet
($r_B{>\atop\sim}1.5$ fm).
On the other hand, for small $m$ the
difference $m_B-m_H\approx a\Lambda_{QCD}$ is dominated by gluonic effects that remain
nonzero even in the chiral limit $m\rightarrow 0$. For $m_B-m_H\approx(200-500)~MeV$, which
may be a realistic range, one finds $r^{-1}_B\approx(500-200)~MeV$. For
the upper value $r^{-1}_B\approx
500MeV$ it seems questionable if one can observe a linearly rising piece of
the heavy quark potential at all, whereas for $r^{-1}_B\approx~200MeV$ this should be
possible. We conclude that it is indeed conceivable that for realistic QCD$_3$ a Yukawa
description (\ref{8.1}) becomes valid. The possible dual
description of the heavy quark potential either by string breaking or by a Yukawa potential
adds another facet to the postulated duality between the Higgs and confinement pictures.

It may be an interesting task for future lattice simulations to find out how the string
breaking distance $r_B$ depends on the quark mass $m$. It seems also worthwhile to
study for which shape of $\hat{\alpha}_s(r)$ in eq. (\ref{8.1}) one can achieve agreement with
phenomenological constraints on the heavy quark potential. In
momentum space the quantity
\begin{equation}
\tilde{\alpha}(q^2)=\alpha_v(q^2)\frac{q^2}{q^2+\bar{M}^2_{\rho}(q^2)}
\end{equation}
should be close to
\begin{equation}
\tilde{\alpha}_R(q^2)=\frac{4\pi}{9}\frac{1}{\ln(1+q^2/\Lambda^2_R)}
\end{equation}
in the momentum range relevant for charmonia. Here $\tilde{\alpha}_R$ corresponds to the
phenomenologically successful Richardson potential \cite{Ri} with $\Lambda_R=400MeV$. One
may also have to take into account that due to the running of the gauge coupling the
effective gluon mass $\bar{M}_{\rho}(q^2)$ will be momentum dependent.

\section{Conclusions}

We have presented a possible Higgs description of the vacuum of two-flavor
QCD$_2$ in terms of $\bar qq$ and $qq$-condensates. Isospin
and baryon number are conserved and electric charges are integer.
The quarks can be associated with baryons and the gluons with
vector mesons. Most strikingly, one finds excitations with
the quantum numbers of strange mesons $(K^*)$ and strange baryons
$(\Sigma,\Lambda)$ in the spectrum of QCD$_2$.

In this picture the gluons acquire a mass by the Higgs mechanism. We find that an
isospin singlet vector meson $(\omega)$ remains massless
in absence of a diquark condensate. This contrasts with
three-flavor QCD$_3$. However, QCD$_2$ admits also a diquark
condensate in the vacuum which can give a mass to
the $\omega$-meson \cite{BerW}.
Despite the spontaneous breaking of the ``standard'' baryon
number $B$ by the diquark condensate, there remains a new conserved
baryon number $B'$. The ``hyperon-like states''
$\Sigma$ and $\Lambda$ are neutral with respect to $B'$.
In addition, nontrivial gluon condensates
are possible in QCD$_2$. All these features differ from
QCD$_3$, and we speculate about
a phase transition from QCD$_3$ to QCD$_2$ as
the strange quark mass increases beyond a critical value.

Two-flavor QCD$_2$ cannot be tested directly by experimental
observation. On the
other hand, QCD$_2$ can be simulated in lattice calculations,
and this may provide important tests for our picture.
The possible tests concern both the properties of the vacuum
and, perhaps at a later stage, the transition to the high density
state. The
most striking signal would be a phase transition in the vacuum
properties as a function of the strange quark mass.
The characteristics of the high density transition in
QCD$_2$ depend crucially on the vacuum properties.

We also have made a first attempt to understand the vacuum
properties of gluodynamics (without light quarks) in terms of
colored gluonic condensates. The partial failure of scalar
condensates to reproduce an acceptable glueball spectrum still leaves open the possibility
that a type of color-spin locking \cite{QCDGS} may offer an interesting alternative.
In this scenario  a residual rotation symmetry is
composed of ordinary rotations accompanied by gauge transformations.
It seems not completely excluded that such a color-spin locking condensate could
also occur in QCD$_2$ and perhaps give a mass to the $\omega$-meson
even in absence of a diquark condensate. As an (perhaps more likely) alternative gluo-
dynamics admits no Higgs picture and spontaneous color breaking occurs only in presence
of sufficiently light quarks. For all the various alternatives
investigated in this note we have found interesting qualitative
changes in the transition from QCD$_3$ to QCD$_2$. We suggest that lattice
simulations could clarify this interesting issue.

\medskip
The author would like to thank J. Berges and U. J. Wiese for collaboration
and stimulating discussions.

\end{document}